\documentclass{aa}
\usepackage{psfig}
\begin{document}

  \thesaurus{02 (11.05.2; 11.04.1; 11.06.1; 11.19.3; 12.05.1)}

   \title{H$\alpha$ emitting galaxies and the cosmic star formation
    rate at z $\simeq$ 2.2. \thanks{Based on observations collected at
    the European Southern Observatories on La Silla and Paranal,
    Chile. Also based on data from the ESO Imaging Survey and HST
    archives.}}


   \author{A.F.M. Moorwood\inst{1}\and P.P. van der Werf\inst{2}\and
   J.G. Cuby\inst{3}\and E. Oliva\inst{4}}
          
   \offprints{A. Moorwood}

   \institute{European Southern Observatory,
              Karl-Schwarzschild-Str. 2, D-85748 Garching, Germany\\ email:
              amoor@eso.org \and Leiden Observatory, P.O Box 9513, NL-2300 RA
              Leiden, The Netherlands\and European Southern
              Observatory, Alonso de Cordova 3107, Santiago, Chile\and
              Osservatorio Astrofisico di Arcetri, Largo E. Fermi 5,
              I-50125 Firenze, Italy}

   \date{Received 3 July 2000/Accepted 9 August 2000}
\authorrunning{Moorwood et al.}
\titlerunning{H$\alpha$ emitting galaxies at z = 2.2}

   \maketitle

   \begin{abstract}

   An infrared imaging survey in narrow band filters around 2.1$\mu$m
has yielded $\sim$ 10 candidate H$\alpha$ emitting galaxies at z
$\simeq$ 2.2 of which 6 have been subsequently confirmed
spectroscopically with ISAAC at the ESO VLT. The survey reached a
limiting line flux of $\sim$ 5x10$^{-17}$ erg cm$^{-2}$ s$^{-1}$ and
covered 100 arcmin$^{2}$ including the Hubble Deep Field South (HDFS)
WFPC2 and STIS fields. This is the largest  spectroscopically confirmed
sample of high redshift galaxies selected by narrow band infrared
imaging. None of the objects falls within the areas of the deep HST
images but some are visible in the WFPC2 flanking fields and the ESO
Imaging Survey (EIS) Deep images of HDFS.  Only one of the objects
observed by HST appears to be an interacting system. Absence of
[NII]$\lambda\lambda$6548,6584 line emission in the spectra is
consistent with them being high ionization and/or low metallicity
systems. The observed velocity dispersions imply masses of typically
10$^{10}$M$\odot$ and a rotation curve obtained for one galaxy yields
an inclination corrected rotational velocity of $\simeq$ 140 km
s$^{-1}$ at 3 kpc which is within the range of nearby disk
galaxies.The absolute B magnitude of this galaxy lies 3 magnitudes
above the local Tully-Fisher relationship.  Star formation rates of
the individual galaxies derived from the H$\alpha$ fluxes are 20-35
M$\odot$ yr$^{-1}$ without any correction for extinction whereas SFRs
derived from the rest frame UV continuum fluxes of the same galaxies
are up to a factor of 4 lower - consistent with lower extinction to
H$\alpha$. Comparison with the HST NICMOS grism H$\alpha$ survey of
Yan et al. (1999) reveals little or no evolution in the H$\alpha$
luminosity function between z $\sim$ 1.3 and 2.2. The inferred star
formation rate density of 0.12 M$\odot$ Mpc$^{-3}$ yr$^{-1}$ is also
equal to that most recently estimated from the UV continuum fluxes of
galaxies at z $\simeq$ 3-4.5 by Steidel et al. (1999). Spectroscopy
covering H$\beta$ and [OIII]$\lambda\lambda$4959,5007 is planned to
gain further insight into the extinction and metal abundance in these
galaxies.

\end{abstract}
\section{Introduction}

Knowledge of both the global star formation history of the Universe
and the nature of individual star forming galaxies at high redshift
are essential to our understanding of galaxy formation and
evolution. Out to z $\simeq$ 1 the star formation rate density is
observed to have increased substantially. The most commonly referenced
study, based on star formation rates inferred from the rest frame UV
continua of CFRS galaxies, yields a factor $\simeq$ 15 increase,
equivalent to luminosity evolution of (1+z)$^{4}$ (Lilly et al. 1996
). At higher redshifts, the major breakthrough came with the detection
of large numbers of the so-called Lyman Break Galaxies at
z$\geq$3. These objects are recognizable in deep UV-visible broadband
images by their absence of continuum flux due to absorption at
wavelengths shorter than the Lyman limit. In a seminal contribution to
the field, Madau et al. (1996) combined star formation rates derived
from the UV continua of these galaxies with those at lower redshifts
to produce a plot of star formation rate density (SFRD) versus z which
implied a decline at $z\geq 3$ relative to $z = 1$ and suggestive of a
possible peak at $z \simeq 2$. The most recent versions of this
diagram which take into account extinction, however, yield both higher
values of the SFRD and suggest that it may actually be rather flat
from $z = 1$ to possibly beyond $z = 4$ (Steidel et al. 1999).
 
We present here the results of a programme whose objectives were to
establish a sample of star forming galaxies at $z\simeq 2$ both for
spectroscopic follow-up studies and to determine the SFRD. Because of the
difficulty of finding and spectroscopically confirming galaxies at
this redshift in the visible we adopted the technique of
narrowband infrared imaging around 2.1$\mu$m to survey for H$\alpha$
line emission at z$\simeq$2.2. As a star formation tracer this line
also has advantages relative to the UV continuum in that it is more
directly related to the youngest hot stars, and hence the current star
formation activity, and is expected to suffer lower extinction. The
extinction can also be estimated by measuring the Balmer
decrement. Spectroscopic measurements of H$\alpha$ in selected CFRS
galaxies (Glazebrook et al. 1999), H$\alpha$ and H$\beta$ in Lyman
Break galaxies (Pettini et al. 1998) and results presented here do in
fact yield SFRs which are a factor of a few higher than derived from
the UV continuum without extinction correction. For spectroscopic
follow-up, narrow band filter imaging also provides the advantage that
the wavelength of the redshifted H$\alpha$ line can be selected to
fall in a clean region between the forest of OH sky lines which
hampers near infrared spectroscopy. The highest redshift accessible is
$\simeq$ 2.5 beyond which the H$\alpha$ line is redshifted out of the
clean part of the K band window and the sensitivity of groundbased
observations falls dramatically due to the increasing thermal
background.

At lower redshifts, the H$\alpha$
luminosity function at $z = 0$ has been measured by Gallego et al.(1995)
and at z $\simeq$0.2 by Tresse \& Maddox (1998). A spectroscopic grism
survey for H$\alpha$ at z = 0.6-1.8 has also been conducted with
NICMOS on the HST and has yielded the H$\alpha$ luminosity function
and SFRD corresponding to a mean redshift of $\simeq$ 1.3 (Yan et
al. 1999). In principle, therefore, it is now possible to trace the
star formation history from z = 0 to 2.5 using H$\alpha$ emission
alone.

Although the advent of large format infrared arrays has made high z
H$\alpha$ surveys feasible, the tradeoff between sensitivity and area
coverage remains a critical issue and one which is dependent on the
scientific aim. The largest area survey to date remains that of
Thompson et al. (1996) who targeted emission at the redshifts of
selected quasars over a total area of 276 arcmin$^{2}$ to a $3\sigma$
flux limit of $\simeq$ 3.5x10$^{-16}$ erg s$^{-1}$ cm$^{-2}$. Only one
candidate object at z = 2.43 was detected and later confirmed
spectroscopically by Beckwith et al. (1998). Several surveys have
subsequently gone deeper over smaller areas and have detected more
candidates but predominantly associated with targeted absorption line
systems which are not representative of the SFRD on large scales
(Mannucci et al. 1998, Teplitz et al. 1998, van der Werf et
al. 2000). These surveys have been mostly conducted in the K band to
target z $\geq$ 2 galaxies and, to our knowledge, none of the
candidates has yet been confirmed spectroscopically. Most recently,
however, deep 2.12$\mu$m imaging of the Hubble Deep Field North has
been used to measure H$\alpha$ in two and
[OIII]$\lambda\lambda$4959,5007in another two galaxies with known
spectroscopic redshifts (Iwamuro et
al. 2000). Their deep 2x2$'$ image reached a $3\sigma$ flux limit of
3.4x10$^{-17}$ erg cm$^{-2}$s$^{-1}$ and the lines are identified as
H$\alpha$ in two of the objects and [OIII]$\lambda$5007 in the other
two.

The project described here started with an infrared imaging survey in
narrow band filters around 2.1$\mu$m conducted with SOFI (Moorwood et
al. 1998) at the ESO NTT telescope.  It reached 3$\sigma$ flux limits
of $\simeq$ 5-12 x 10$^{-17}$ erg cm$^{-2}$ s$^{-1}$ over a total area
of $\simeq$ 100 sq. arcmin including the WFPC2 and STIS fields in the
Hubble Deep Field South (Williams et al. 2000). Apart from the slightly higher
redshift quasar in the STIS field there are no known redshift
'markers' close to our target redshift and we therefore believe our
results to be representative of the field galaxy population at
$z\simeq2.2$. Spectroscopic confirmation of most of the best candidate
emission line objects obtained subsequently with ISAAC at the VLT
(Moorwood et al. 1999) has demonstrated the validity of the survey
technique and also provided additional insight into the nature of the
galaxies detected.

We describe here both the imaging and spectroscopic
observations in Sect. 2; present the results, together with additional
groundbased and HST data, in Sect. 3; discuss the possible nature of
these objects and derive the star formation rate density at z = 2.2 in
Sect. 4 and summarize our conclusions in Sect. 5.

Mainly for ease of comparison with published results we have adopted a
cosmology with H$_{0}$ = 50 km s$^{-1}$ Mpc$^{-1}$ and q$_{0}$ = 0.5
throughout.


\section{Observations}
\subsection{Infrared imaging survey}
SOFI, the infrared imager/spectrometer at the ESO NTT telescope
(Moorwood et al. 1998) was used in August 1998 to conduct the
narrow-band filter search for H$\alpha$ emitting galaxies at z
$\simeq$ 2.2. This instrument is equipped with a 1024x1024 pixel
Rockwell 'Hawaii' array detector which covers a field of view of
nearly 5x5 arcmin on the sky with pixels of 0.29$''$ in its large
field imaging mode. Three survey fields of this size were selected -
one each centred on (but larger than) the WFPC2 and STIS fields in the
Hubble Deep Field South and one on an anonymous field about
$30^{o}$ away which was selected to be devoid of bright objects on the
DSS. The nominal J2000 field centers were WFPC2: 22 32 56.2, -60 33
02.7; STIS: 22 33 37.7, -60 33 29 and Blank: 20 50 00, -67 50 00. All
three fields were observed in both a 1\% FWHM filter centred at
2.09$\mu$m, in a region of low OH background emission, and the
broad-band Ks (2.16$\mu$m) filter. The STIS and WFPC2 fields were
additionally observed in a 1.3\% filter centred at 2.12$\mu$m.

All images were taken using the `autojitter' mode with the telescope
being offset by random amounts of up to 20$\arcsec$ between individual
exposures of typically 6x30s and 6x10s in the narrow and broadband Ks
filters respectively. Total exposures were 4-6hrs in the narrow and
$\simeq$ 1hr in the Ks  filter yielding 3
$\sigma$ line flux detection limits of 5-12x10$^{-17}$ erg
s$^{-1}$ cm$^{-2}$ in the different fields. The seeing was close to  1$''$ for
all the observations. Allowing for losses at the field
edges due to the jitter technique our survey covered 40 sq. arcmin
(4200 Mpc$^{3}$ co-moving) at z = 2.24 and 60 sq. arcmin (4500
Mpc$^{3}$ co-moving) at z = 2.18. Table~\ref{tab.fields} summarizes the flux
limits, areas and volumes surveyed in the SOFI fields.

\begin{table}
      \caption[]{The SOFI survey fields} \label{tab.fields}
         \[ \begin{array}{lllllll} \hline \noalign{\smallskip}
{\rm Field} & \lambda(\mu m) & {\rm z}
&\Delta z&{\rm Flux} ^{a}&{\rm Area}^{b}&{\rm Vol}^{c}  \\ 
 \\ \noalign{\smallskip} \hline
         \noalign{\smallskip}
 
{\rm WFPC2} &2.09&2.18&0.03&4.8&18.9&1426\\ {\rm WFPC2}&
2.12&2.23&0.043& 6.2&20.3&2194 \\ {\rm STIS}&2.09&2.18&0.03&9&18&1356
\\{\rm STIS}&2.12&2.23&0.043&8&19&2055\\{\rm
Blank}&2.09&2.18&0.03&12&20.6&1557\\

   \noalign{\smallskip} \hline \end{array} \]
\begin{list}{}{}
\item[$^{\mathrm{a}}$] 3$\sigma$ flux limit in units of 10$^{-17}$erg
cm$^{-2}$s$^{-1}$
\item[$^{\mathrm{b}}$]   in arcmin$^{2}$
\item[$^{\mathrm{c}}$]   co-moving Mpc$^{3}$
\end{list}
   \end{table}

\subsection{ISAAC spectroscopy}

\begin{figure}[htb]
\psfig{figure=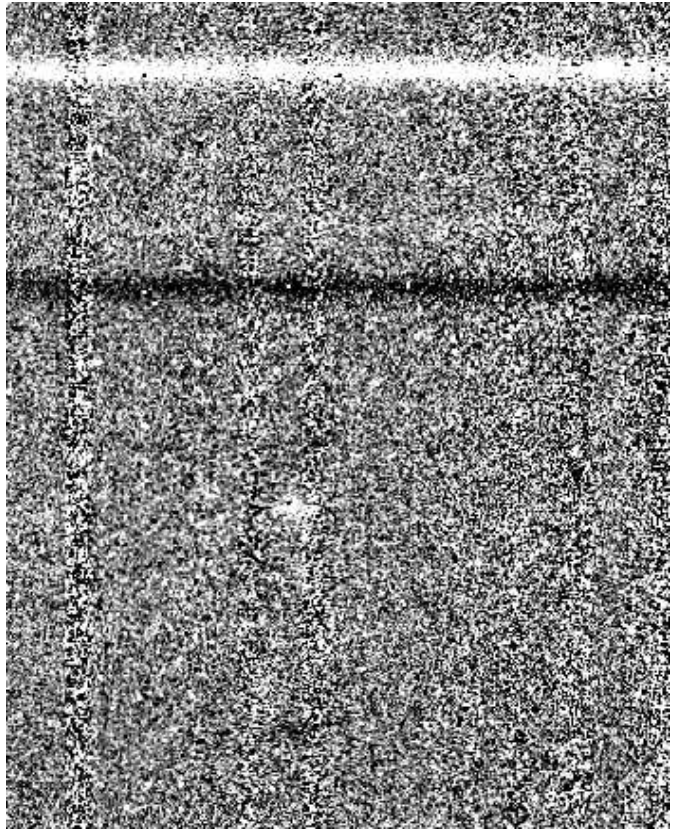,width=8.8cm}
\caption{Sky subtracted 2D ISAAC spectrum of 338.165-60.518 and the
reference galaxy above. Emission is only detected at the position of
the H$\alpha$ line in the programme galaxy whereas the reference
galaxy is much brighter in the continuum. Note the positive and 2
negative images due to the sky subtraction technique used and the
increased shot noise at the position of the OH sky lines.}
\label{fig.184spec}
\end{figure}

Infrared spectra around 2.1$\mu$m of all 8 candidate H$\alpha$
emitters detected at $\geq$ 3$\sigma$ plus 3 with s/n in the range 2-3
were obtained with ISAAC (Moorwood et al. 1999) at the ESO VLT in
June/July 1999. Observations were made using the SW Rockwell 1024x1024
pixel Hawaii array and the medium resolution grating whose resolving
power x slit width product is about 2500 at the wavelengths
observed. As the seeing was typically only around 1-1.5$''$ the 2$''$
slit was in fact used in all cases except for the simultaneous
observation of two galaxies with the 1$''$ slit when the seeing was
around 0.6$''$. The target galaxies were acquired using the imaging
mode of ISAAC. Because of their faintness, advantage was taken of the
long (2$'$) slit to accurately centre the targets by angular
offsetting relative to two nearby brighter objects in the field. In
each case, the first step was to centre the two reference objects in
the slit. Using the telescope rotator, the field was then rotated by
the measured angular offset of the target galaxy relative to the line
between the two reference objects in the SOFI images to avoid errors
due to uncertainties in the exact scale. The telescope was then offset
to centre one of the reference objects as well as the target galaxy in
the slit. This technique facilitates location of the faint object
spectrum in the 2D sky subtracted frames and provides a check on the
telescope tracking and flexure during the observations. In two cases,
it was actually possible to centre two target objects plus a reference
simultaneously in the slit. Each observation comprised four 15min
on-chip integrations with the object moved between exposures by
5-10$''$ along the slit in an ABBA
sequence. Fig.~\ref{fig.184spec} shows the reduced 2D, sky
subtracted, spectrum of 338.165-60.518 plus the reference galaxy above
obtained by spatially shifting the A-B and B-A frames by the offset
and then averaging. This yields a positive object spectrum and 2
negative ones of half the amplitude on either side. This technique not
only yields good sky subtraction but also allows faint lines to be
distinguished from cosmetic detector effects which only yield a
positive and one negative spectrum.1D spectral traces integrated over
the spatial extent of the objects were extracted using standard MIDAS
routines. Flux calibration has been derived from observations of the
standard star HD216009 made with both the 1$''$ and 2$''$ slits which
yielded closely similar fluxes.

\begin{figure}[htb]
\vspace{0cm}
\hspace{0cm}
\psfig{figure=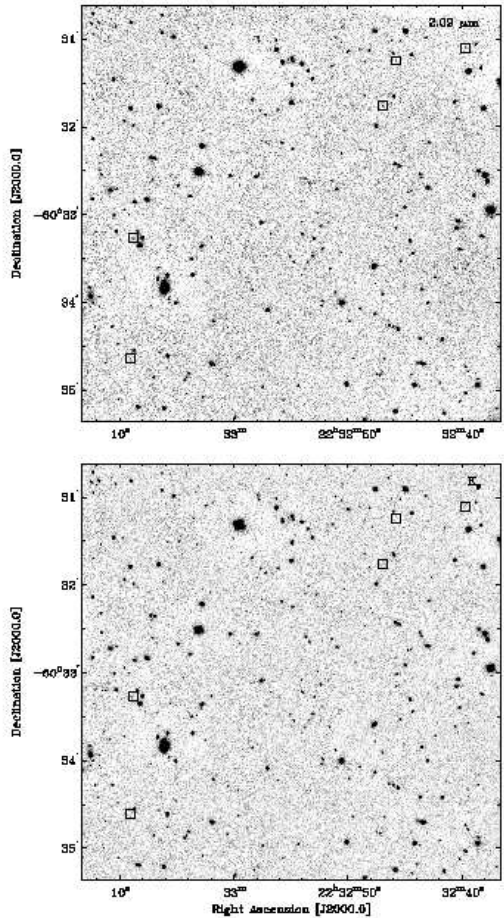,width=12cm}
\vspace{-1cm}
\caption{2.09$\mu$m narrow (upper panel) and Ks broad band band
images centred on the WFPC2 field. Field is 5x5$'$ with N at the top
and E to the left. The squares identifiy the H$\alpha$ emitting
candidates confirmed spectroscopically.}
\label{fig.wfpcnbbb}
\end{figure}

\section{Results}
\subsection{SOFI survey}
\begin{figure}[htb]
\psfig{figure=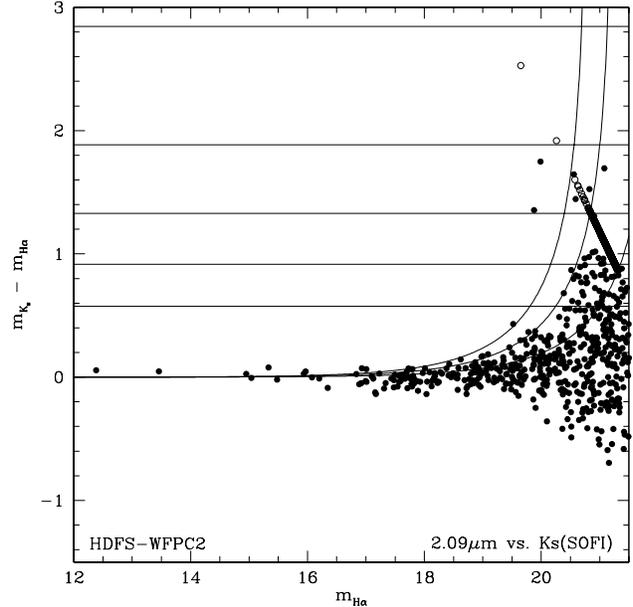,width=8.8cm}
\caption{m(Ks)-m(H$\alpha$) vs m(H$\alpha$) for the 2.09$\mu$m survey
centred on the WFPC2 HDFS field. The solid lines represent the 1,2 and
3$\sigma$ line detection limits and the horizontal lines correspond to
rest equivalent widths of 50, 100, 200 500\AA\, and infinity. The open
symbols represent sources only detected in the narrow band filter and
are therefore lower limits.}
\label{fig.trumpet}
\end{figure}

 Fig.~\ref{fig.wfpcnbbb} shows the narrow 2.09$\mu$m (upper) and Ks
broad band SOFI images centred on the WFPC2 field as an
example. Squares identify the H$\alpha$ emitting candidates
subsequently confirmed spectroscopically. Catalogues of the objects
in all 5 survey fields were made using SExtractor (Bertin \& Arnouts
1996).Candidate line emitting objects were
then selected on the basis of their excess narrow versus broad band
flux in plots of m$_{nb}$-m$_k$ vs m$_{nb}$. Fig.~\ref{fig.trumpet} is
the plot for the WFPC 2.09$\mu$m field showing the loci for flux
excess at the 1,2 and 3$\sigma$ level as well as the limits on
equivalent width. Due to the area coverage and depth reached the large
number of galaxies detected in total means that the {\it trumpet} shaped
region occupied by non-line emitting galaxies is well defined by the
data. Several candidates exhibiting excess emission are clearly
visible. Most of the candidates in fact were found in this field due
to the lower flux limit achieved relative to the others and/or
possibly clustering effects.

Although the 2.09$\mu$m filter corresponds to H$\alpha$ at the highest
redshift of several absorption systems along the line of sight to the
STIS quasar no candidates were actually detected in this filter/field
combination implying that our data are not affected by clustering
associated with this object.

Table~\ref{tab.cand} lists the candidates detected at $\geq$ 3$\sigma$
plus those with lower s/n ratios for which ISAAC spectra were
obtained. Also given are the Ks magnitudes and H$\alpha$ line fluxes
measured both in the narrow band filter and 
from the spectra. Agreement between the photometric and spectroscopic
line fluxes is actually excellent if account is taken of the fact that
the largest discrepancies are due to those lines which did not fall
close to the central wavelength of the narrow band filter.

\begin{table}
      \caption[]{SOFI candidate/confirmed H$\alpha$ emitters}
         \label{tab.cand} \[ \begin{array}{llllll} \hline
         \noalign{\smallskip} {\rm Field} & \lambda(\mu m) & {\rm
         Object} & {\rm Ks} & {\rm s/n}^{a} & {\rm H}\alpha ^{b} \\
         \noalign{\smallskip} \hline \noalign{\smallskip} {\rm WFPC} &
         2.09 & 338.165-60.518 & 21.7&5 & 7.8/9\\ & &
         338.191-60.521&-&4&7.6/9\\ & & 338.193-60.527 & - & 2.8 &
         4.4/ \leq5\\ & & 338.196-60.529 & 21.2 & 5 & 7.7/10\\ & &
         338.287-60.555 & 20.7& 5 & 8.1/9\\ & & 338.288-60.577 &
         21.4 & 2.1 & 3.3/8\\ & & 338.290-60.572 & 21.3 & 2.6 &
         4.6/\leq5\\ 

 {\rm WFPC}&2.12 & 338.300-60.539 &-& 3.2 & 7.3/\leq5\\ {\rm
         STIS}&2.12& 338.366-60.547 & 21.2& 5 & 14.6/13\\ & &
         338.382-60.523 & 20.5 & 3.4 & 8.9/\leq5\\
	& & 338.407-60.558^{c} & 14.8& 600& 960/-\\
  {\rm Blank} & 2.09 & 306.300-67.863 & -&5& 26/\leq5\\
	
         \noalign{\smallskip} \hline \end{array} \]
\begin{list}{}{}
\item[$^{\mathrm{a}}$] detection significance in imaging survey
\item[$^{\mathrm{b}}$] 10$^{-17}$erg cm$^{-2}$s$^{-1}$ in NB filter/spectra
\item[$^{\mathrm{c}}$] quasar in STIS field
\end{list}
   \end{table}

\begin{figure*}[htb]
\psfig{figure=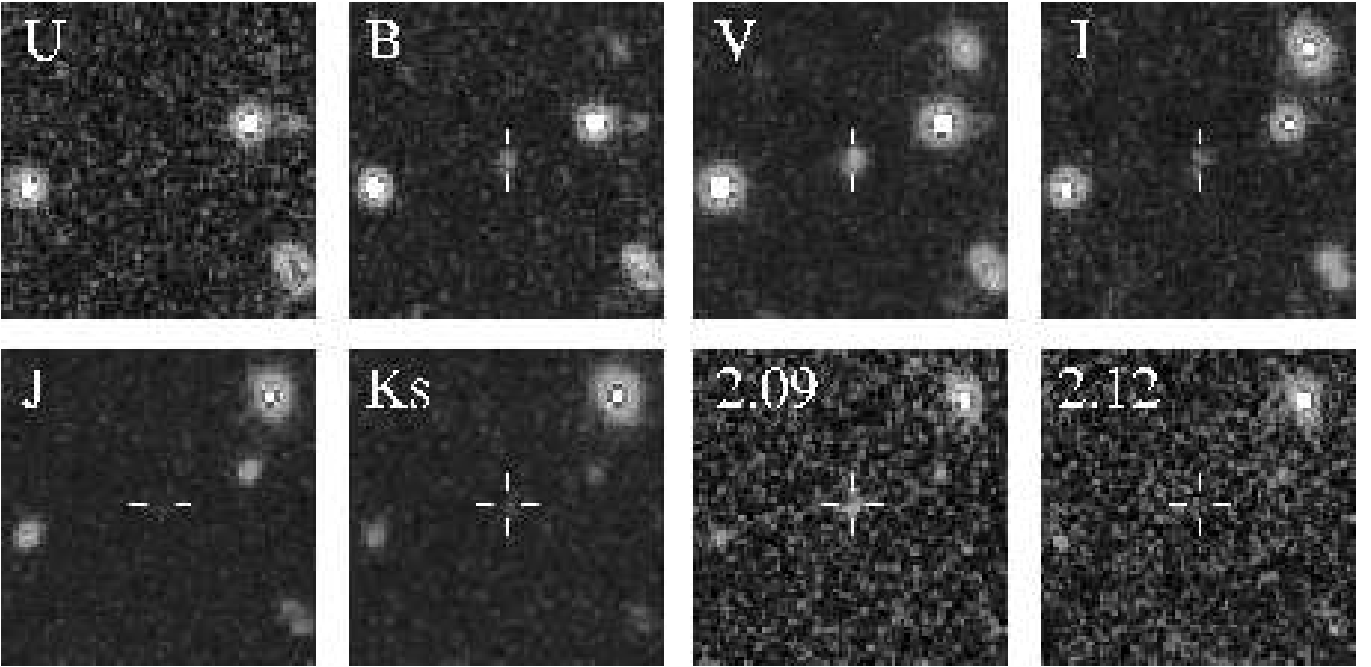,width=18cm}
\caption{Images of a 20x20$''$ region centred on 338.196-60.529. The
U,B,V,I images were obtained in the EIS Deep programme and the J, Ks
NB 2.09$\mu$m and NB 2.12$\mu$m images in the present work. Note that
the object at the centre is well detected in B to I but not U and,
relative to the other objects in the field, is much brighter in the
NB2.09$\mu$m than the Ks filter and undetected in the NB 21.12$\mu$m
filter.  This appearance is characteristic of a galaxy at z=2.18
suffering Lyman forest absorption in the U band and with a strong,
redshifted H$\alpha$ emission line falling within the NB 2.09$\mu$m filter.}
\label{fig.cand}
\end{figure*}

\subsection{Colours and morphology}
 Most, but not all, of our candidates fall in the ESO SUSI2/SOFI EIS
Deep fields (http://www.eso.org/science/eis/). Complete UBVRIJHKs
photometry is available for a few and the most interesting result is
that 3 of the H$\alpha$ emitting galaxies show extremely red U-B
colours ($\geq$2 mag.) which are presumably due to Lyman forest
absorption at the survey redshift of 2.2. This provides support for
the H$\alpha$ detections but also implies that these objects could
have been detected in a photometric redshift survey. Images of
338.196-60.529 are shown in Fig.\ref{fig.cand}. Note that the galaxy is
visible in the B to I bands but not U and is bright in the NB
2.09$\mu$m filter whereas it is not detected in the NB 2.12$\mu$m
filter and only barely in Ks.

\begin{figure}[tb]
\psfig{figure=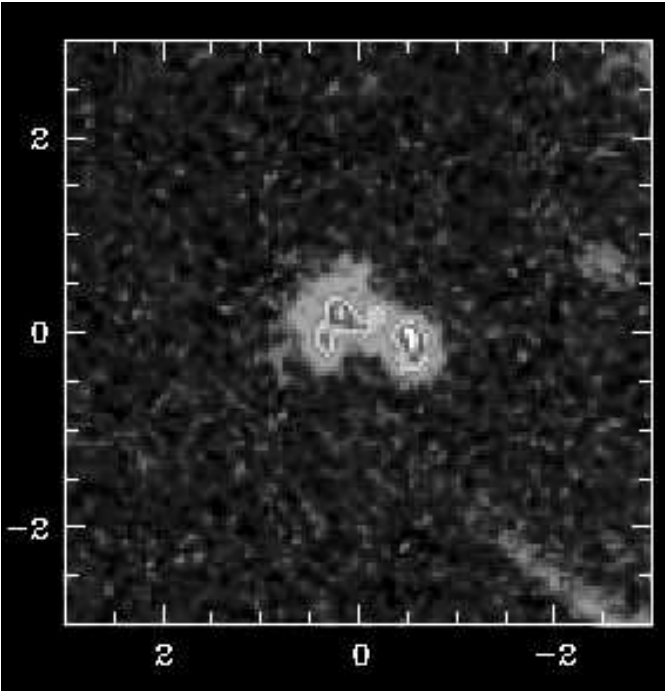,width=8.8cm,angle=0}
\caption{WFPC2 HDFS flanking field F814W image of
338.287-60.555. Scale is in arcsec..}
\label{fig.918}
\end{figure}

\begin{figure}[tb]
\psfig{figure=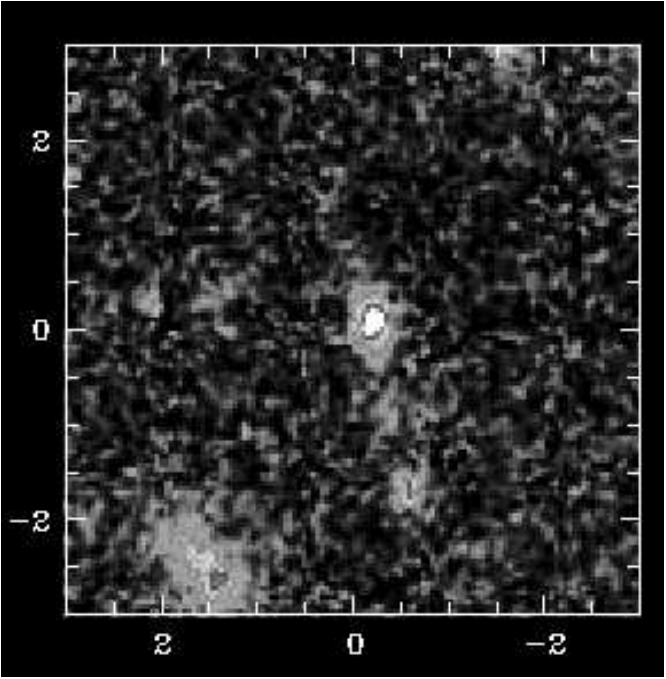,width=8.8cm,angle=0}
\caption{WFPC2 HDFS flanking field F814W image of
338.288-60.577. Scale is in arcsec..}
\label{fig.1375}
\end{figure}

All of the candidate line emitting objects in our 5x5$'$ images fall
outside the smaller deep HDFS fields observed with WFPC2 and STIS
although some were observed in the WFPC2 flanking fields
(http://www.stsci.edu/ftp/science/hdfsouth/hdfs.html). The F814W image
of 338.287-60.555 shown in Fig.~\ref{fig.918} is of particular
interest and suggests that this is an interacting system with 2 or 3
components within $\simeq$ 1$''$ (8 kpc). This object is also the most
diffuse of the sample in the infrared narrow band and spectral
images. It appears to be the only such case within our sample.  The
F814W image of 338.288-60.577 shown in Fig.~\ref{fig.1375} is also of
interest as this object is the only one in which we see a rotation
curve in H$\alpha$. The HST image shows a single galaxy whose major
axis is roughly N-S in which case it is within about 10$^{\rm o}$ of the
slit orientation used. The other objects for which F814W images are
available, 338.165-60.518, 338.191-60.521,338.196-60.529 show no
particularly interesting morphological structure but their fluxes
allow the comparison made below of SFRs derived from the rest frame UV
continuum and H$\alpha$.


\subsection{ISAAC spectroscopy}

\begin{figure*}[htb]
\psfig{figure=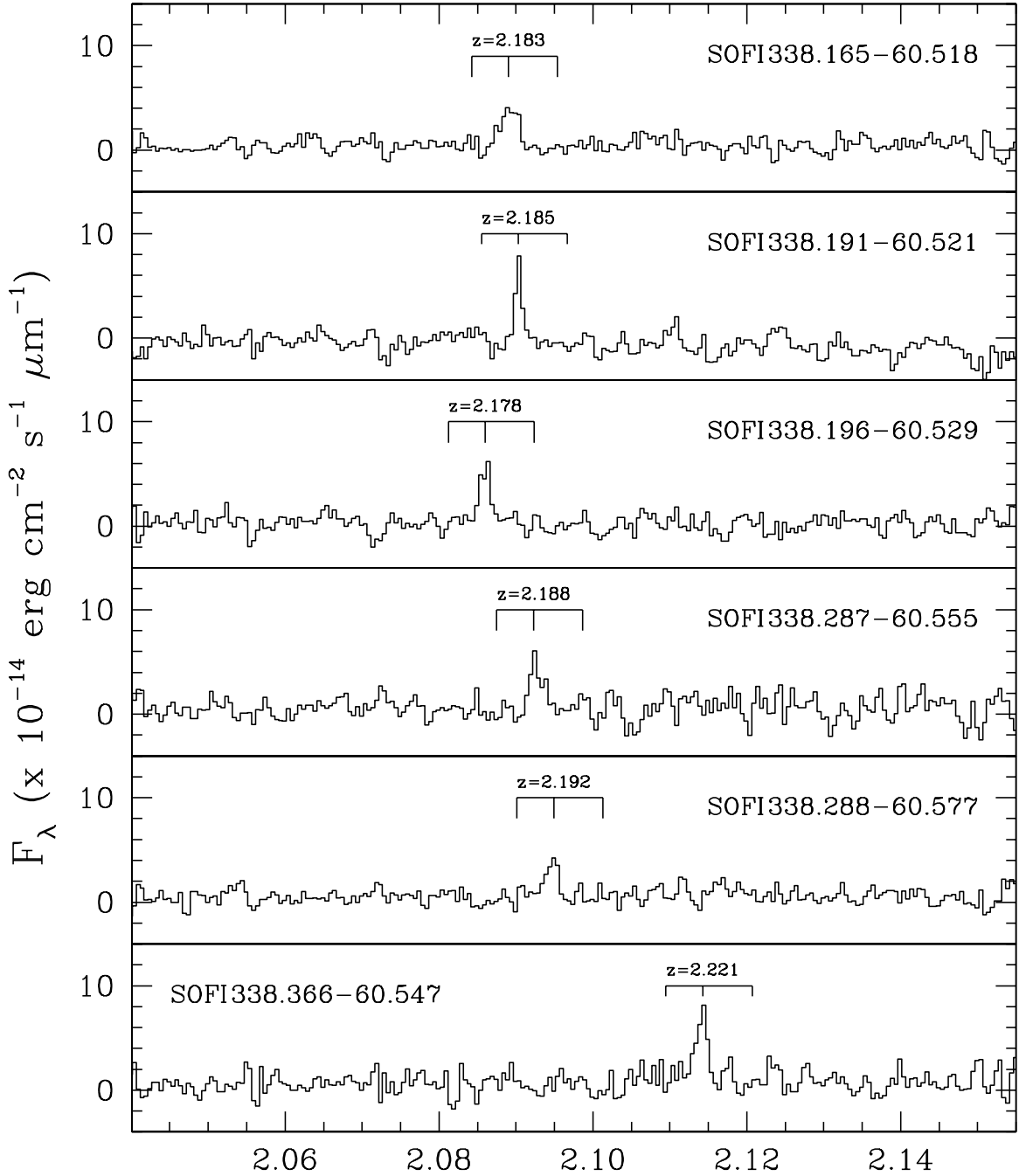,width=18cm}
\caption{ISAAC spectra of  H$\alpha$ emitting galaxies. The
tick marks under the redshift labels show the expected positions of the
[NII](6548,6584\AA) lines assuming the detected line is H$\alpha$.}
\label{fig.spectra}
\end{figure*}

Spectra of the 6 objects (the STIS quasar has been excluded) showing a
clear emission line at the expected wavelength are shown in
Fig.~\ref{fig.spectra}.  The ratio of spectroscopically confirmed to
total H$\alpha$ candidates as a function of their s/n in the imaging
survey are 5/6 at $\geq$ 4$\sigma$, 5/8 at $\geq$3$\sigma$; and 1/3 at
$\leq$ 3$\sigma$. In the last group, although only detected at
2$\sigma$ in the survey, the spectroscopically measured flux of
338.288-60.577 turned out to be higher because its redshift places the
line away from the centre of the NB filter. This object was also
observed under the best seeing conditions ($\simeq$0.6$''$) with the
1$''$ slit and is the only one which exhibits a clear rotation
curve. For the other two s/n $\leq$ 3 sources it appears that the flux
sensitivity reached in the spectra might have been insufficient even
if these objects are real. This illustrates the difficulty of
establishing completeness at the survey limit. Non confirmation of
306.300-67.536, the brightest of those detected at $\geq$ 4$\sigma$ is
potentially the most surprising except that it was not detected in the
Ks filter and therefore may be spurious.  In the case of
non-detections there also remain the possibilities of poor centering
or that the line is coincident in position with an atmospheric OH
line.

The first and most important conclusion from the spectra is that most
of the objects detected at $\geq$4$\sigma$ in our imaging survey do
actually exhibit an emission line at the correct wavelength. It is
important to stress this as very few of the infrared objects detected
by this technique previously have been spectroscopically confirmed. In
principle, a problem of identification remains as only a single line
appears in the spectra. On the other hand, this eliminates the most
serious alternative to H$\alpha$ which is
[OIII]$\lambda\lambda$4959,5007. The 4959/5007 ratio of the [OIII]
doublet is 0.33 so both lines should  be visible at the s/n achieved.The
only other serious possibility is [OII]$\lambda$3727 but this is both
intrinsically fainter than H$\alpha$ and the objects would have to be
at $z = 4.6$. This is also a doublet with almost equally intense
components although they would only be marginally resolvable in our
spectra.  A marginal detection of redhifted [OIII]$\lambda$5007
obtained in a short H band spectrum of 338.366-60.547 is also
consistent with the line at 2.1$\mu$m being H$\alpha$.  

If, as appears most likely, the detected lines are all H$\alpha$ then
it is of interest that [NII]$\lambda\lambda$6548,6584, whose expected
positions are marked on Fig.~\ref{fig.spectra} are not detected. The
[NII]/H$\alpha$ ratio shows wide variations with galaxy type,
ionization degree, abundance etc. The most commonly used diagnostic
diagram for emission line galaxies is the plot of
[OIII]$\lambda$5007/H$\beta$ versus
[NII]$\lambda$6584/H$\alpha$. Based on this diagram in Gallego et
al.(1997), low [NII]$\lambda$6584/H$\alpha$ ratios are characteristic
of high ionization, low metallicity systems. For a sample of Lyman
Break galaxies at z $\sim$ 3 observed with NIRSPEC and ISAAC (Pettini
2000) the [OIII]$\lambda$5007/H$\beta$ ratios are typically $\geq$ 3
which corresponds to the same part of the diagram and would imply
[NII]$\lambda$6584/H$\alpha$ $\leq$ 0.1 if these objects are of
similar nature. Absence of the [NII] lines in our relatively low s/n
spectra is therefore not particularly surprising and can actually be
taken as evidence for the high ionization degree and/or low metal
abundances which is to be expected at this redshift.
\begin{table*}[htb]
      \caption[]{Derived quantities from ISAAC spectra.}
         \label{tab.spectra} \[ \begin{array}{llccccccc} \hline
         \noalign{\smallskip} {\rm Field} & {\rm Source} & {\rm z} &
         {\rm Slit}('')& {\rm H}\alpha^{a} & {\rm L}({\rm
         H}\alpha)^{b}&{\rm SFR}({\rm M}_\odot/yr)&{\rm
         FWHM(obs.)}^{c}&\sigma_{v}^{c}\\ \noalign{\smallskip} \hline
         \noalign{\smallskip}
         {\rm WFPC209}&338.165-60.518&2.183&2&9\pm0.5&3.066&24.2&479&175\\
         &338.191-60.521&2.185&2&9\pm1.2&3.07&24.3&151&-\\
         &338.196-60.529&2.178&2&10\pm1&3.39&26.8&202&-\\
         &338.287-60.555&2.188&2&9\pm0.7&3.08&24.3&350&\geq108\\
         &338.288-60.577&2.192&1&8\pm0.5&2.75&21.74&301&117\\
         {\rm STIS}212&338.366-60.547&2.221&2&13\pm1.7&4.61&36.42&273&\geq50\\

         \noalign{\smallskip} \hline \end{array} \]
\begin{list}{}{}
\item[$^{\mathrm{a}}$] 10$^{-17}$erg cm$^{-2}$s$^{-1}$
\item[$^{\mathrm{b}}$] 10$^{42}$erg s$^{-1}$
\item[$^{\mathrm{c}}$] km s$^{-1}$
\end{list}
   \end{table*}

\begin{table}
      \caption[]{Comparison of SFRs from H$\alpha$ and UV continuum}
         \label{tab.sfr} \[ \begin{array}{llllllc} \hline \noalign{\smallskip}
         {\rm Source}& {\rm I}^{a}& {\rm L}_{uv}^{b}& {\rm SFR}_{uv}^{c}
         &{\rm SFR}_{{\rm H}\alpha}^{c}&{\rm R}({\rm H}_{\alpha}/uv)\\
         \noalign{\smallskip} \hline \noalign{\smallskip}
338.191-60.521&2({\rm E})&4.6&6.5&24.3&3.7\\
338.196-60.529&2.8({\rm W})&6.5&9.1&26.8&3.0\\
338.287-60.555&10.3({\rm W})&23.8&33.3&24.3&0.7\\
338.288-60.577&5.2({\rm W})&11.9&16.7&21.74&1.3\\
338.366-60.547&3.1({\rm E})&20&28&36&1.3\\

         \noalign{\smallskip} \hline \end{array} \]
\begin{list}{}{}
\item[$^{\mathrm{a}}$] 10$^{-19}$erg cm$^{-2}$s$^{-1}$$\AA^{-1}$
(E-EIS, W-WFPC2)
\item[$^{\mathrm{b}}$] 10$^{28}$ erg s$^{-1}$Hz$^{-1}$ at $\simeq$ 2500\AA
\item[$^{\mathrm{c}}$] M$\odot$/yr
\end{list}
   \end{table}

\section{Discussion}
\subsection{Star formation rates}
Star formation rates for the spectroscopically confirmed galaxies have
 been computed using the formula SFR(M$\odot$/yr) = 7.9x10$^{-42}$
 L(H$\alpha$)(erg s$^{-1}$) from Kennicutt(1998) which is appropriate
 for continuous star formation and a Salpeter IMF extending from
 0.1-100M$\odot$. The values of 20-35 M$\odot$/yr, reported in
 Table~\ref{tab.spectra}, are higher than found in the disks of late
 type spirals but lower than in the most extreme nearby starburst
 galaxies. As no extinction correction has been applied to the
 H$\alpha$ fluxes, however, the actual values could be higher. The
 canonical value is A$_{H\alpha}$ = 1.1 mag for nearby spirals
 (Kennicutt 1983) but increases to $\simeq$ 5-40 mag. in Ultraluminous
 Infrared Galaxies which are believed to harbour the most powerful
 starbursts (Genzel et al. 1998). Unfortunately, spectra around
 H$\beta$ which would allow an estimate of the extinction from the
 Balmer decrement could not be obtained within the available telescope
 time. For some of our objects, however, we have been able to extract
 I band fluxes from HST (WFPC2 F814W) observations of the HDFS
 flanking fields and EIS Deep images obtained with the ESO NTT. The
 flux conversions used were as given in the respective file headers.
 As the wavelength corresponds to $\simeq$ 2500$\AA$ in the rest frame
 it is possible for these galaxies to compute the SFR also from their
 rest frame UV continua using the analogous formula SFR =
 1.4x10$^{-28}$ L(1500-2800$\AA$)(erg s$^{-1}$Hz$^{-1}$) from
 Kennicutt(1998) which is based on the same model assumptions. This is
 done in Table~\ref{tab.sfr} where it can be seen that the SFRs
 deduced from H$\alpha$ are higher on average by a factor $\simeq
 2$. For the two H$\alpha$ emitting objects detected in HDFN, Iwamuro
 et al. (2000) also derive SFRs which are almost a factor of 2 higher
 than those estimated from the UV continuum. Although there are
 considerable uncertainties in these numbers they do support the
 expectation, based on the wavelength dependence of dust extinction,
 that extinction to the UV continuum is higher than to H$\alpha$. They
 are also roughly consistent with the extinction correction at this
 redshift most recently adopted in deriving the SFRD from the UV
 continua of Lyman break galaxies (Steidel et al. 1999). It is worth
 noting, however, that the H$\alpha$ emission could actually suffer
 higher extinction if the youngest stars are still more heavily
 enshrouded in dust than those responsible for the bulk of the UV
 continuum. This appears not to be the case.  The fact that these
 galaxies are detected in the rest frame UV continuum also argues
 against very high absolute extinction values although the possibility
 remains that a substantial fraction of the star formation activity
 could be obscured to both the UV continuum and H$\alpha$.

\subsection{Dynamics}

Despite the fact that most spectra were obtained with a 2$''$ slit
(FWHM = 213 km s$^{-1}$) and the relatively low s/n ratios some of the
detected emission lines are resolved and their estimated velocity
dispersions given in Table~\ref{tab.spectra} range up to $\simeq$
200km s$^{-1}$ with a mean value $\simeq$ 80 km s$^{-1}$. This value
is similar to that found from infrared spectroscopy of Lyman Break
galaxies (Pettini 2000). Applying the formula in Devereux et
al. (1987) the implied mass within the central few kpc is typically
$\sim$ 10$^{10}$M$\odot$. This of course assumes that the line widths
are related to the mass and not due to the kinematics of the gas e.g
winds.

\begin{figure}[htb]
\psfig{figure=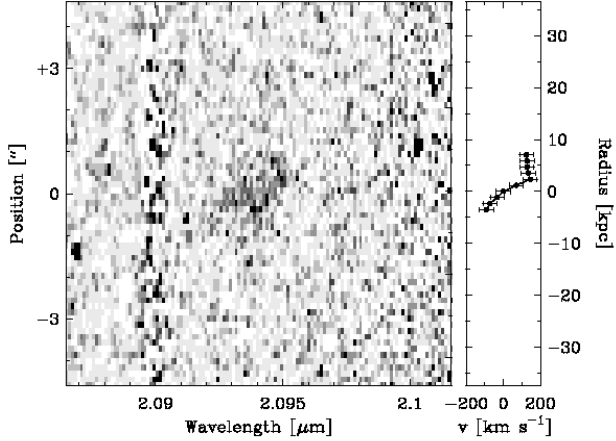,width=11cm,angle=-90}
\caption{The left panel is the 2D spectrum of 338.288-60.57 whose line
emission is 'tilted' relative to the dispersion direction due to the
galaxy rotation. The right panel is the rotation curve obtained by
fitting Gaussians to the spectra at each spatial position along the
slit. }
\label{fig.rot}
\end{figure}

The clearest evidence that we are actually observing relatively
massive systems is provided by the observations of 338.288-60.577
which were made with the 1$''$ slit when the seeing was $\leq$
0.6$''$. Its HST I band image shown in Fig.~\ref{fig.1375} shows this
galaxy to be extended $\simeq$ 1$''$ in the N-S direction and aligned
within $\simeq$10$^{\rm o}$ with the slit. Assuming an effective
extent of 6kpc along the major axis the mass implied by the measured
velocity dispersion of 117 km s$^{-1}$ is $\simeq$ 2 x
10$^{10}$M$\odot$. Of more significance in this case, however, is the
clear tilt of the H$\alpha$ line which we attribute to ordered
rotation. Although the s/n ratio is low, this is clearly evident in
the 2D spectral image which is shown in Fig.~\ref{fig.rot} together
with the corresponding rotation curve obtained by fitting Gaussian
profiles to the emission at each spatial position along the slit. The
fact that the line is tilted is robust. It appears in spectra reduced
from each half of the data set independently and in a separate
spectrum obtained with the 2$''$ slit on another night. The details of
the extracted rotation curve are clearly somewhat uncertain due to the
low s/n ratio. In particular, the reality of the flattening observed
on the positive velocity side would need to be confirmed by higher s/n
observations. The basic information of interest, however, is the observed p-p
velocity spread of 247$\pm 30$ km s$^{-1}$ over a distance of 0.6$''$
or $\simeq$ 6kpc. The intrinsic value increases to 275$\pm 30$ km
s$^{-1}$ after correction for the inclination of i = 64$\pm 5^{\rm o}$
deduced from the aspect ratio a/b = 2.15$\pm 0.3$ in the I band image after
correction for the PSF. Our observations thus imply a rotational
velocity of 138$\pm 15$ km s$^{-1}$ at r $\simeq$ 3 kpc which may also
be the terminal velocity if the observed flattening is real. This is
comparable to what is seen in nearby disk galaxies whose 21cm HI
rotation curves tend to flatten at velocities in the range 100-300km
s$^{-1}$ at radii of 1-5kpc (Begeman et al. 1991). On this evidence,
therefore, it appears that well developed, massive systems  were already
in place at z $\simeq$ 2.

338.288-60.577 was also observed with SOFI in the EIS deep survey and
has an apparent H band magnitude of 24.34 (AB) which corresponds to a
rest frame absolute B magnitude of M$_{B}$ = -22.4. For its FWHM
H$\alpha$ velocity of $\simeq$240 km s$^{-1}$ this is about 3
magnitudes brighter than expected for a nearby galaxy falling on the
Tully-Fisher relation (unless the full rotation curve extends over
$\simeq$ 1000 km s$^{-1}$ and only flattens at a radius $\geq$ 12 kpc
which is highly unlikely). A similar result has been obtained for
Lyman Break galaxies at z $\simeq$ 3 (Pettini 2000). Qualitatively,
it is of course to be expected that these highly active star forming
galaxies exhibit enhanced B luminosity to mass ratios. It is
nevertheless interesting that these first quantitative estimates yield
values similar to the total increase in the SFRD out to these
redshifts.

\subsection{H$\alpha$ Luminosity Function}

Gallego et al. (1995) have shown that the H$\alpha$ luminosity function of
galaxies in the local universe is well fitted by a Schechter function
of the form
\begin{equation}
\phi(L)dL
=\phi^{*}(L/L^{*})^{\alpha}e^{-L/L^{*}}d(L/L^{*})
\end{equation}

with $\alpha$ = $-$1.3, $\phi^{*}$ = 6.3 x10$^{-4}$Mpc$^{-3}$ and L$^{*}$
= 1.4 x10$^{42}$ erg s$^{-1}$.

They computed the volume density of galaxies $\Phi$(log L) per
Mpc$^{3}$ per 0.4 interval of log L(H$\alpha$) where
\begin{equation}
\Phi(log L)( d log L)/0.4 = \phi(L)dL
\end{equation}
At higher redshifts in the range z $\simeq$ 0.6-1.8, Yan et al. (1999)
find that the H$\alpha$ luminosity function of galaxies detected in
their HST NICMOS grism survey is also well fitted with a function
having the same form but with
\begin{equation}
 \phi^{*} = 1.7{\rm x}10^{-3}{\rm Mpc}^{-3}\,
{\rm and}\, L^{*} = 7 {\rm x}10^{42} {\rm erg}{\rm s}^{-1}
\end{equation}

Between $z= 0$ and $\simeq$ 1 therefore the density of H$\alpha$
emitting galaxies increases by a factor 2.7 and L$^{*}$(H$\alpha$) by
a factor 5. The total star formation rate density thus increases by a
factor of 13.5. Although this is comparable to that deduced from the
UV continua of CFRS galaxies (Lilly et al. 1996) the true ratio must
actually be larger as the Gallego et al. results have been corrected
for extinction derived from the Balmer decrements whereas those of Yan
et al. have not.

For comparison we have estimated the co-moving volume density of
H$\alpha$ emitting galaxies found at z = 2.2 in our survey. For each
candidate galaxy in Table ~\ref{tab.cand} we have computed the maximum
co-moving volume V$_{max}$ in which it could have been detected by
summing the survey volumes for which the required flux sensitivity was
reached. Within luminosity bins of d log L(H$\alpha$) = 0.4 we then
computed
\begin{equation}
\Phi(log L) = \sum1/V_{max}
\end{equation}

For each bin the statistical errors were also computed as the square
roots of the variance i.e the sum of the squares of the inverse
volumes. 

\begin{figure}[htb]
\psfig{figure=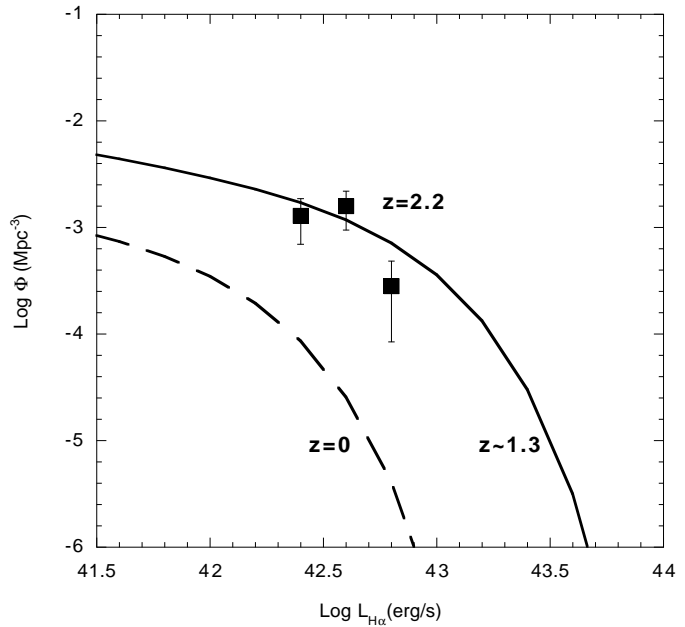,width=9cm}
\caption{H$\alpha$ luminosity functions. The filled squares are from
this work and correspond to the assumption that all survey candidates
detected at $\geq$ 3 $\sigma$ are real.  The curves are the best fit
Schechter functions to the data of Yan et al. (1999) at z $\sim$
1(solid) and those of Gallego et al. (1995) at z=0 (dashed). }
\label{fig.lum}
\end{figure}

The results are shown in Fig.~\ref{fig.lum} where the filled squares
are from the present work and the curves are the Schechter function
fits to the luminosity functions at z=0 and z$\simeq$ 1.3 given by
Gallego et al. (1995) and Yan et al. (1999). 

The candidates found in our
survey occupy a relatively limited range in H$\alpha$ luminosity -
limited at the lower end by sensitivity and at the higher by area
coverage. The narrow band imaging technique used to find our sources
is also subject to an equivalent width threshold which increases with
decreasing source flux. Estimated from the photometry, the actual rest
frame EWs of our detected sources are in the range 50 - 700\AA\,
whereas about 30\% of the galaxies in the Gallego sample have EWs
$\leq$ 50\AA\,. At the high luminosities detectable in our survey,
however, the EW distribution can be expected to be shifted to higher
values in which case this selection effect is expected to have a
relatively small effect. 

It is of interest that our value for the
highest luminosity bin is somewhat low relative to the z $\sim$ 1.3
curve and could indicate a deficiency in extremely high star formation
rate galaxies. As discussed above, such an effect may be expected if
the extinction increases with SFR as found in nearby starburst
galaxies. The density for the lowest luminosity bin is probably too
low due to incompleteness.  The nominal co-moving survey volumes given in
Table~\ref{tab.fields} are for a redshift range corresponding to the
FWHM of the narrow band filters. As the filters have closer to
Gaussian than rectangular shapes, however, the flux limits depend on
redshift within the passband.  It is not possible to correct the
imaging data directly for this because the line wavelengths and hence
true fluxes are not known {\it a priori}. Estimates of this effect
have been made, therefore, using the measured transmission
curves of the narrow-band filters. For the 2.09$\mu$m filter the
nominal $\Delta$z corresponding to the FWHM is 0.03. For the flux
limit reached in the WFPC field the effective $\Delta$z for 3$\sigma$
detections decreases from 0.038 to 0.027 for objects with log
L(H$\alpha$) = 43 to 42.4 but falls to 0.0085 at log L(H$\alpha$) =
42.2. For the 2.12$\mu$m filter, with a nominal $\Delta$z =0.043, the
effective values on the same field decrease from 0.06 to 0.027 in the
range log L(H$\alpha$) = 43 to 42.4 and is essentially 0 at 42.2. For
these fields therefore the conclusion is that errors in the volume
densities are relatively small for log L(H$\alpha$) $\geq$ 42.4 but
can be large at lower luminosities.  

As not all candidates in the survey have been spectroscopically
confirmed it is also possible that the plotted points are actually too
high. As a check, therefore, we have computed separately the volume
density of the spectroscopically confirmed candidates in the WFPC2.09
field.  As all of these objects have log L(H$\alpha$) in the range
42.4-42.6, the effective co-moving volume expected is very close to
the nominal one adopted based simply on the FWHM of the NB filter. As
objects with log L(H$\alpha$) = 42.4 are brighter than the flux limit
in this field the detections at this luminosity should also be
complete. In fact the volume density of log $\Phi$ = -2.5
(+0.17,-0.25) obtained is a factor 2 higher than the value of -2.8
(+0.14,-.22) derived from the complete imaging survey. This is
actually not surprising given that this field is the deepest and
appears to contain a small group or cluster. Despite that, the
magnitude of the effect is actually only at the level of the quoted
statistical uncertainties. Within these uncertainties, our overall
conclusion is that the comparison of our result with that of Yan et
al. is consistent with either no or only modest evolution in the
H$\alpha$ luminosity function between $z\simeq 1.3$ and 2.2.

\subsection{Star Formation Rate Density}

The total H$\alpha$ luminosity density at z= 2.2 can be estimated by
integrating the Schechter function
\begin{equation}
 L_{tot}(H\alpha) = \int\phi(L)LdL
 = \phi^{*}L^{*}\Gamma(2+\alpha) 
= 1.3 \phi^{*}L^{*}
\end{equation}

 As we cannot improve on the fit provided by the curve of Yan et al. we
use this to obtain L$_{tot}$(H$\alpha$) = 1.55x10$^{40}$ erg
s$^{-1}$. To convert this to a star formation rate density we then use
the relation SFR(M$_{\odot}$yr$^{-1}$) =
7.9x10$^{-42}$L(H$\alpha$)(erg s$^{-1}$) from Kennicutt(1998) which
yields a SFRD = 0.12 M$_{\odot}$yr$^{-1}$. This is, of course,
identical to the Yan et al. value at z $\sim$1 and the implied
flatness of the SFRD vs z curve in this range is independent of
extinction as no correction has been applied in either case. It is
of interest to note that our value of the SFRD is also almost
identical to that derived by Steidel et al.(1999) from extinction
corrected UV continuum measurements of galaxies at z $\sim$3-4.5. This
direct comparison is not strictly fair as Steidel et al. (1999) imposed
a cut-off at 0.1 L$^{*}$ which could lead to a factor $\simeq$ 2
underestimate in the total SFRD. On the other hand, correction for
extinction would also lead to some increase in our value. It is
important to stress here, however, that agreement at this level is
remarkable enough given the small size of our sample, issues of
completeness and the large uncertainties in the shape of the adopted luminosity
function. 

\section{Conclusions}
\begin{itemize}
\item  A 2.1$\mu$m narrow band imaging survey conducted with SOFI at the ESO
NTT has yielded about 10 candidate H$\alpha$ emitting galaxies with
fluxes down to a few x 10$^{-17}$ erg cm$^{-2}$s$^{-1}$ over an area
of 100 arcmin$^{2}$ which includes the HDFS WFPC2 and STIS fields.

\item Based on HST WFPC2 observations of the HDFS flanking fields only
one of these objects appears to be an interacting system with 3
components within $\sim$ 10kpc. Three objects appearing in EIS Deep
images of the HDFS exhibit extremely red U-B colours, consistent with Lyman
forest absorption at the target redshift of z = 2.2

\item Six of the best candidates have been confirmed spectroscopically
using ISAAC at the ESO VLT. Although only a single emission line is
seen in each case its only plausible identification is
H$\alpha$. Absence of the [NII]$\lambda\lambda$6548,6584 lines is
consistent with the high [OIII]/H$\beta$ ratios observed on higher
redshift Lyman Break galaxies and indicative of high ionization and/or
low metallicity systems. This is the largest sample of
spectroscopically confirmed, high redshift, galaxies selected by
narrow band infrared imaging.

\item Star formation rates derived from the H$\alpha$ fluxes are in
the range 20-35M$\odot$/yr without extinction correction and are, on
average, a factor $\sim$ 2 higher than those derived from the UV
continua of the same galaxies.

\item The velocity dispersions $\sim$ 100km s$^{-1}$ are similar
to those measured in Lyman Break galaxies and imply masses
$\sim$10$^{10}$M$\odot$ provided they are related to mass and not
winds. More direct evidence that these are relatively well developed
systems is provided by a rotation curve obtained for one galaxy which
yields a rotational velocity of $\simeq$ 140km s$^{-1}$ at a radius of
3 kpc which is comparable with nearby disk galaxies. The absolute B
magnitude of this galaxy is $\simeq$ 3 magnitudes brighter than
expected from the local Tully-Fisher relationship.
\item Although sampling only a narrow range around log L(H$\alpha$)
$\simeq$42.6, comparison of our data with the results of the H$\alpha$
NICMOS grism survey conducted by Yan et al. (1999) imply little or no
evolution in the H$\alpha$ luminosity function and hence of the Star
Formation Rate Density between z $\sim$1.3 and 2.2. 

\item Our best estimate of 0.12 M$\odot$yr$^{-1}$Mpc$^{-3}$ for the
SFRD at z =2.2 is, within the statistical uncertainties, equal to that
derived from the UV continuum flux of Lyman Break galaxies at z =
3-4.5 by Steidel et al. (1999).

\item Additional spectroscopy covering H$\beta$ and
[OIII]$\lambda$4959,5007 is now planned in order to measure the extinction
and estimate the metal abundances in these systems.

\end{itemize}

\begin{acknowledgements}

We are grateful to Max Pettini and Lin Yan for helpful discussions.
     
\end{acknowledgements}

\end{document}